# Achievable information rates estimates in optically-amplified transmission systems using nonlinearity compensation and probabilistic shaping

Daniel Semrau[†], Tianhua Xu[†,*], Nikita A. Shevchenko, Milen Paskov, Alex Alvarado, Robert I. Killey, Polina Bayvel

*Optical Networks Group, Department of Electronic and Electrical Engineering, University College London (UCL), London, UK, WC1E 7JE*
*\*Corresponding author: tianhua.xu@ucl.ac.uk*

**Achievable information rates (AIRs) of wideband optical communication systems using ~40 nm (~5 THz) EDFA and ~100 nm (~12.5 THz) distributed Raman amplification are estimated based on a first-order perturbation analysis. The AIRs of each individual channel have been evaluated for DP-64QAM, DP-256QAM, and DP-1024QAM modulation formats. The impact of full-field nonlinear compensation (FF-NLC) and probabilistically shaped constellations using a Maxwell-Boltzmann distribution were studied and compared to electronic dispersion compensation. It is found that a probabilistically shaped DP-1024QAM constellation combined with FF-NLC yields AIRs of ~75 Tbit/s for the EDFA scheme and ~223 Tbit/s for the Raman amplification scheme over 2000 km standard single mode fibre transmission. © 2017 Optical Society of America**

*OCIS codes: (060.0060) Fiber optics and optical communications; (060.1660) Coherent communications; (060.4370) Nonlinear optics, fibers.*

http://dx.doi.org/10.1364/OL.42.000121

In the past four decades the data rates of optical communication systems have experienced an astonishing increase from 100 Mbit/s per fiber in the 1970s to 10 Tbit/s in current commercial systems. The technical milestones that unlocked the feasibility of these high rates were wavelength division multiplexing (WDM), improved fiber design and fabrication, optical amplifications and coherent detection [1]. Erbium-doped fiber amplifiers (EDFAs) and Raman fiber amplifiers have made it possible to extend the usable fiber bandwidth in the past, however these amplification technologies are now viewed as limiting the accessible optical spectrum to ~5 THz and ~10-15 THz, respectively [2].

Higher information rates are vital to cope with the growing data demand. In coded transmission systems, achievable information rates (AIRs) are an important figure of merit as they correspond to the net data rates that can be achieved by transceiver based on soft-decision decoding [Sec. IV, 3]. For a given bandwidth, higher AIRs can be achieved by using closer channel spacing (e.g. Nyquist spacing) and higher modulation formats. However, in such systems, AIRs are limited by fiber nonlinear distortion arising from the optical Kerr effect. Much research has been devoted to improve this limit through a variety of nonlinear compensation (NLC) techniques, e.g., digital back-propagation, optical phase conjugation, nonlinear Fourier transform, twin-wave phase conjugation, and others, as reviewed in [4]. AIRs can also be increased by means of signal shaping [5-7]. Geometric shaping enhances the AIR by changing the uniform constellation grid, while probabilistic shaping (PS) uses a non-uniform probability distribution on the constellation points. In this Letter we consider probabilistic shaping as it offers several advantages over geometric shaping as discussed in [Sec I, 5].

Investigations of AIRs for C-band and beyond rely on the analytical calculations since the computational complexity of split-step simulations become infeasible. In previous work, AIRs were evaluated for an EDFA system with a total bandwidth of 5 THz [8] considering electronic dispersion compensation (EDC) only. In addition, AIRs have been investigated for a bandwidth of 4.3 THz [9] assuming ideally distributed Raman amplification and EDC only. In all the reported work, the NL distortion coefficient of the central channel has been employed as the NL distortion coefficients for all channels over the entire optical bandwidth to estimate the AIRs [8-10]. However, those coefficients are smaller in outer channels due to lower inter-channel nonlinearities.

In our previous work [11], the AIRs of Nyquist-spaced optical communication systems were analytically estimated using a large-bandwidth first-order perturbation analysis for ~100 nm (~12.5 THz) backward pumped Raman amplification over a standard single mode fiber (SSMF), with the applications of EDC and full-field nonlinear compensation (FF-NLC).

In this Letter, the results of [11] are extended to include ~40 nm (~5 THz) EDFA amplification, higher-order modulation formats and probabilistically shaped constellations, to demonstrate the possibility of approaching the fundamental limits of the two most considered amplification schemes. The nonlinearities within each channel are individually evaluated and AIRs are investigated with the applications of EDC and FF-NLC. In addition, it is shown that the use of PS together with FF-NLC can further improve the AIRs and approach the limits imposed by the interactions between signal and amplified spontaneous emission (ASE) noise. Using dual-polarization 1024-ary quadrature amplitude modulation (DP-1024QAM) and a transmission distance of 2000 km (25×80 km), AIRs of ~75 Tbit/s in the EDFA amplified system and ~223 Tbit/s in the Raman amplified system can be achieved with FF-NLC and PS. These results highlight the potential AIRs of optical communication systems in the nonlinear regime.

Additionally, it is shown that using FF-NLC and PS, DP-256QAM system can achieve the same AIRs as the DP-1024QAM system, when the transmission distance exceeds 3200 km in the EDFA scheme and 6000 km in Raman amplification scheme.

In the following, equations are presented to assess performance of EDFA and Raman amplified communication systems as basis of AIR calculations. It is assumed that the NL distortion, often called NL interference noise, is approximately Gaussian and additive. In the regime of weak non-linearity (around optimum launch power), the NL distortion in dispersion-unmanaged long-haul transmission systems can be considered to be Gaussian and independent from other noise sources. This assumption has been validated for EDFA and Raman amplified links [8-10]. Therefore, the signal-to-noise ratio (SNR) at the receiver is then given by

$$\text{SNR} = \frac{P}{P_\text{n} + P_\text{s-s} + P_\text{s-n}}, \quad (1)$$

with the launch power $P$, the ASE noise $P_\text{n}$, the NL interference noise arising from signal-signal interactions $P_\text{s-s}$ and the NL interference noise arising from signal-noise interactions $P_\text{s-n}$. The latter two quantities are modeled as $P_\text{s-s} = N_\text{s}^{1+\varepsilon} \eta P^3$ and $P_\text{s-n} \approx 3\xi\eta P_\text{n}P^2$ with $\xi \triangleq \sum_{k=1}^{N_\text{s}} k^{1+\varepsilon}$, and the NL distortion coefficient $\eta$ is approximately equal for signal-signal interactions and signal-noise interactions [12], as the ASE noise is typically modeled with a Gaussian distribution and the signal is also assumed to be Gaussian. The number of spans $N_\text{s}$ and the factor $\xi$ are responsible for the accumulation of noise along the link and $\varepsilon$ denotes the coherence factor taken from [Eq. (18), 8].

Typical fibre values (SSMF) are used to evaluate the system performance with the parameters shown in Table 1. The ASE noise at the receiver in Nyquist WDM systems are calculated according to [Sec. 6.1.3, 2] for EDFA and [Eq. (6), 13] for Raman amplification.

Given its high accuracy in long-haul, highly dispersive systems with dense modulation formats, a first-order perturbation analysis was used to compute the NL distortion coefficients [3,8-10,12,14]. The assumption of Gaussian signal distributions overestimates the impact of nonlinear distortions with respect to a uniform QAM constellation. In fact, this holds in for most probabilistically shaped QAM constellations, and thus, a Gaussian distribution lower bounds the SNR for most probability mass functions, including Maxwell-Boltzmann. Therefore, all calculated AIRs reported in this letter are guaranteed to be achievable.

The NL distortion coefficient for channel $k$ is calculated with

$$\eta(k) = \frac{1}{R_\text{s}} \int_{\left(\frac{2k-1}{2}\right) \cdot R_\text{s}}^{\left(\frac{2k+1}{2}\right) \cdot R_\text{s}} S(f)\ df, \quad (2)$$

$$S(f) = \frac{16\gamma^2}{27R_S^2} \iint_{-\frac{B}{2}}^{\frac{B}{2}} \Pi\left(\frac{f_1+f_2-f}{B}\right) \rho(f_1,f_2,f)\ \chi(f_1,f_2,f)\ df_1 df_2, \quad (3)$$

with the total bandwidth $B \triangleq N_\text{ch} \cdot \Delta f$ and $\Pi(x)$ denoting the rectangular function. The phased-array factor $\chi(f_1, f_2, f)$ is responsible for noise accumulation over multiple spans and is given by

$$\chi(f_1,f_2,f) = \frac{\sin^2\{2N_s\pi^2(f_1-f)(f_2-f)[\beta_2+\pi\beta_3(f_1+f_2)]\cdot L_s\}}{\sin^2\{2\pi^2(f_1-f)(f_2-f)[\beta_2+\pi\beta_3(f_1+f_2)]\cdot L_s\}}. \quad (4)$$

**Tab. 1:** System parameter values

| Parameters | Values |
|---|---|
| EDFA, Num. of channels ($N_\text{ch}$) | 501 |
| Raman, Num. of channels ($N_\text{ch}$) | 1251 |
| Symbol rate ($R_S$) | 10-Gbaud |
| Channel spacing ($\Delta f$) | 10-GHz |
| Roll-off factor | 0% |
| Total Raman pump power ($P_\text{p0}$) | 5×680 mW |
| Noise figure ($F_\text{n}$) | 4.5 dB |
| Span length ($L_\text{s}$) | 80 km |
| Fiber loss ($\alpha$) | 0.20 dB/km |
| Fiber loss for Raman pump ($\alpha_\text{p}$) | 0.25 dB/km |
| Dispersion ($D$) | 17.0 ps/nm/km |
| Dispersion slope ($S$) | 0.067 ps/nm²/km |
| NL coefficient ($\gamma$) | 1.20 1/W/km |
| Linewidth of transmitter | 0 Hz |
| Linewidth of local oscillator | 0 Hz |

It should be noted that the phased-array factor Eq. (4) is only used to calculate the coherence factors. The four-wave mixing (FWM) efficiency factor $\rho(f_1, f_2, f)$ depends on the amplification scheme. For EDFA amplification and backward pumped Raman amplification the FWM efficiency factors are given by [8,10]

$$\rho_\text{EDFA}(f_1,f_2,f) = \left|\frac{1-e^{-\alpha L_s}e^{j4\pi^2(f_1-f)(f_2-f)[\beta_2+\pi\beta_3(f_1+f_2)]\cdot L_s}}{\alpha-j4\pi^2(f_1-f)(f_2-f)[\beta_2+\pi\beta_3(f_1+f_2)]}\right|^2, \quad (5)$$

$$\rho_\text{Raman}(f_1,f_2,f) = \left|e^{-\frac{C_R P_{P0}}{\alpha}}\int_0^{L_s} e^{-\alpha z}e^{\frac{C_R P_{P0}}{\alpha_\text{p}}e^{\alpha_\text{p} z}}e^{j4\pi^2(f_1-f)(f_2-f)[\beta_2+\pi\beta_3(f_1+f_2)]\cdot z}dz\right|^2, \quad (6)$$

with the Raman gain coefficient $C_\text{R}$ and $j \triangleq \sqrt{-1}$. Here a depolarized pump source and no polarization dependence of the Raman gain are assumed. In reality the Raman gain is polarization-dependent; however, high gain efficiencies can still be realized with the method described in [15].

To reach faster convergence in numerical integrations, the small impact of $\beta_3$ is neglected in the remainder of this Letter. The whole spectrum of NL distortion coefficients is calculated for both discussed amplification schemes and is shown in Fig. 1.

Additionally, the SNR according to Eq. (1) is plotted for EDC only ($P_\text{s-n} = 0$) and for FF-NLC ($P_\text{s-s} = 0$) at optimum and uniform launch power across the spectrum. The SNR at optimum launch power scales as $\Delta\text{SNR}_\text{EDC}\ [\text{dB}] \approx -\frac{1}{3}\eta\ [\text{dB}]$ for EDC only and as $\Delta\text{SNR}_\text{NLC}\ [\text{dB}] \approx -\frac{1}{2}\eta\ [\text{dB}]$ for FF-NLC. Therefore, the change in

SNR is slightly different at the edges of the spectrum in the FF-NLC case compared to the EDC only case. For the central channel Raman amplification outperforms EDFA amplification by 3.2 dB and 4.9 dB for EDC only and FF-NLC, respectively. The bigger margin using FF-NLC is easily understood by looking at the scaling of the FF-NLC gain which is approximately given by $\Delta\text{SNR}_{\text{NLC-gain}} [\text{dB}] \approx -\frac{1}{6}\eta [\text{dB}] - \frac{1}{3}P_n[\text{dB}]$. This relation yields a ΔSNR of 1.63 dB in favour of Raman amplification due to its lower ASE noise contribution.

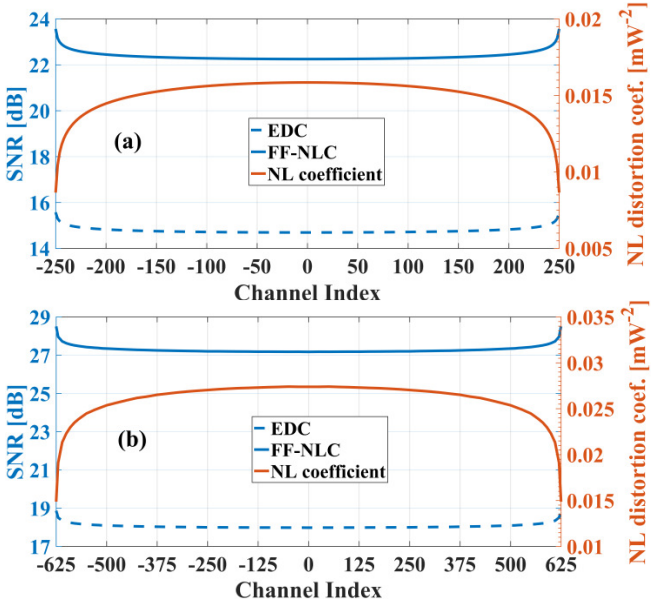

**Fig. 1.** NL distortion coefficients and SNR values at 2000 km (25×80 km). (a) EDFA system (b) Raman amplified system.

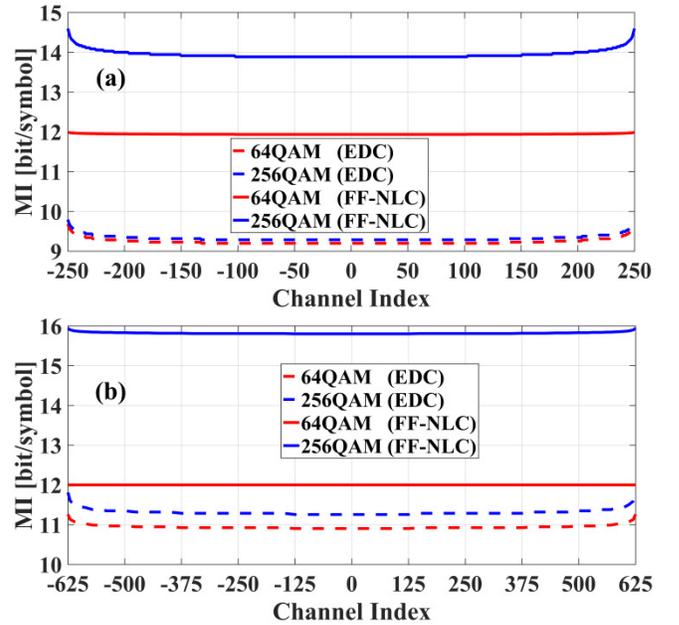

**Fig. 2.** Mutual information for each channel at 2000 km (25×80 km). (a) EDFA system, (b) Raman amplified system.

To compute the AIR, the soft-decision mutual information (MI) for each channel was numerically calculated using Gauss-Hermit quadrature. Here, the MI is a transmission rate that can be achieved, assuming a Gaussian channel law. Fig. 2 shows the results for both amplification schemes.

For the transmission length of 2000 km and EDC-only case, the use of DP-64QAM is a good compromise between the performance and complexity as DP-256QAM gives marginal improvement. However, when FF-NLC is applied the MI for DP-64QAM is saturated at 12 bit/symbol. Higher modulation formats need to be applied such as DP-256QAM as it would yield a MI of ~14 bit/symbol for the EDFA scheme and ~16 bit/symbol for the Raman scheme.

To further increase the MI, a probabilistically shaped constellation was used. The idea is to transmit constellation points with a non-uniform probability according to a certain probability mass function. More information on PS and typical constellation diagrams can be found in [5,6]. In the case of Gaussian noise channel and an average power constrain it can be shown that signal shaping yields a gain of up to 1.53 dB which is referred to as the ultimate shaping gain [Sec. IV-B, 6]. As the optimum input distribution for the nonlinear fiber-optic channel is still under debate, a Maxwell-Boltzmann distribution was applied, since it has been proved to be optimal for QAM constellations in a Gaussian channel [7]. This is consistent with the first-order perturbation analysis of nonlinear distortions.

An important property of nonlinear channels is the SNR dependence on the input distribution. However, as mentioned above, the SNR calculated based on a Gaussian distribution represents a lower bound. To obtain a tighter lower bound, modulation format dependent models can be considered [16,17]. In the EDFA case a closed-form expression is available that approximately corrects for the modulation dependence of cross-channel interference [16]. Applying the parameters used in this Letter, the expression yields a higher optimal SNR of 0.63 dB for DP-64QAM. For higher modulation formats this difference will be slightly smaller. These small corrections (strictly positive) are neglected as the modulation format dependent model will introduce unmanageable additional complexities.

In the following both EDFA and Raman amplification schemes are presented in terms of AIRs for 3 cases, namely EDC-only, FF-NLC, and PS combined with FF-NLC. The results are shown in Fig. 3. A signal-ASE noise interaction limit is shown by calculating $\sum_k 2R_s \log_2[1 + \text{SNR}_{\text{FF-NLC}}(k)]$ assuming a Gaussian channel law and a Gaussian constellation. A limit only considering ASE noise is also plotted by computing $\sum_k 2R_s \log_2[1 + \text{SNR}_{\text{ASE}}(k)]$.

The use of FF-NLC significantly increases the AIRs justifying the current research efforts in nonlinear compensation techniques. Additionally, the AIRs of the Maxwell-Boltzmann shaped input distribution approaches the signal-ASE noise interaction limit for longer distances. For both EDFA and Raman amplified systems, the increase in modulation formats gives a considerable improvement in the AIRs, when the transmission distance is less than 2000 km. For DP-1024QAM and a total transmission distance of 2000 km, the application of FF-NLC can realise AIRs of ~70 Tbit/s and ~215 Tbit/s for the EDFA and the Raman amplification schemes, respectively. This can be pushed further to the signal-ASE noise interaction limit by using PS that yields AIRs of ~75 Tbit/s and ~223 Tbit/s, respectively.

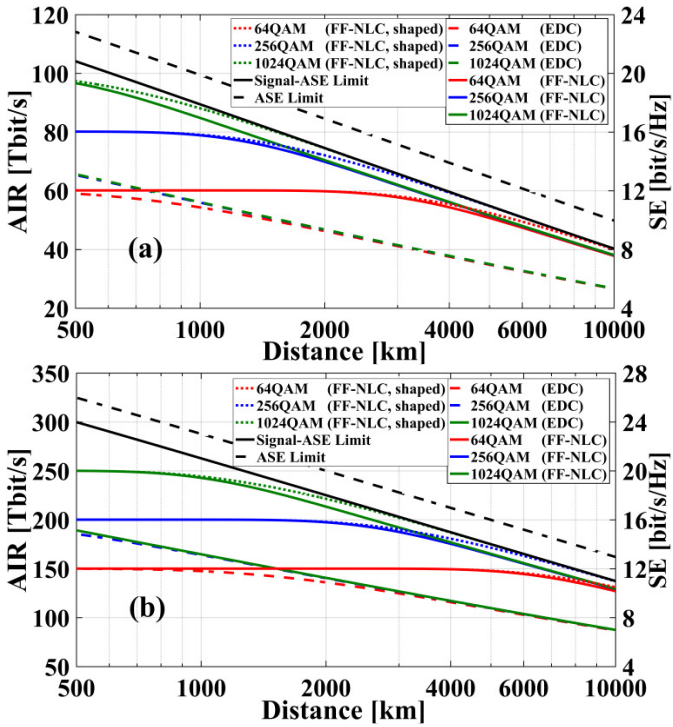

**Fig. 3.** AIRs of optical transmission systems. (a) EDFA system, (b) Raman amplified system.

In addition, it was found that, DP-256QAM can achieve the same AIRs as DP-1024QAM (both in the signal-ASE noise limit) with the use of FF-NLC and PS, when the transmission distance exceeds 3200 km in the EDFA system and 6000 km in the Raman amplified system. This implies that DP-256QAM is sufficient over these long-haul transmission distances.

The model in this work has been applied to a recent experimental record reporting 49.3 Tbit/s transmission over 9100 km using C-band + L-band EDFA [18]. According to our estimation and neglecting the ~4 nm gap between C-band and L-band, this record reaches ~76% of the theoretical AIR for DP-16QAM and EDC only, due to practical limitations. However, overcoming these practical limitations, the use of DP-256QAM, FF-NLC and PS, would potentially double the reported transmission rate and achieve the signal-ASE interaction limit within the same bandwidth.

It is noted that phase noise (PN) contributions of both transmitter and local oscillator were neglected. In practical transmission systems, the PN will interact with dispersion compensation modules in both linear and nonlinear compensation schemes, which may lead to equalization enhanced phase noise (EEPN) [19,20]. The additional impact from EEPN will be the subject of subsequent study.

In conclusion, the achievable information rates for ~40 nm (~5 THz) EDFA and ~100 nm (~12.5 THz) Raman amplified optical communication systems have been analytically investigated in standard single mode fibre transmission. Using a first-order perturbation analysis, the nonlinear distortions have been evaluated in each individual channel. It was shown that the use of DP-1024QAM, wideband nonlinearity compensation and probabilistic shaping can achieve information rates of ~75 Tbit/s and ~223 Tbit/s over 2000 km for the EDFA and the Raman amplification schemes, respectively.

**Funding.** UK EPSRC project UNLOC (EP/J017582/1)

**Acknowledgment.** We thank Prof. A. D. Ellis (Aston University) for insightful discussions, and Tobias Fehenberger (Technical University of Munich) & Gabriel Saavedra (University College London) for helpful contributions.

†These authors contributed equally to this work.